\documentclass[aps,prb,showpacs,floatfix,twocolumn,superscriptaddress]{revtex4-2}
\usepackage{placeins}
\usepackage{flafter}
\usepackage{fontspec} 
\usepackage{newunicodechar}
\newunicodechar{−}{-} 
\newunicodechar{–}{-} 
\usepackage{graphicx}
\usepackage{dcolumn}
\usepackage{graphicx}
\usepackage{float}
\usepackage{relsize}
\usepackage{cancel}
\usepackage{bm}
\usepackage{tikz}
\usepackage{siunitx}
\usepackage{tablefootnote}
\usepackage{physics}

\usepackage[%
pdfauthor={Benedikt Placke},%
pdfstartview=FitH,%
breaklinks=true,%
bookmarks=true,%
colorlinks=true,%
anchorcolor=black,%
citecolor=red,
filecolor=black,%
menucolor=black,%
urlcolor=blue,%
linkcolor=red,%
]{hyperref}

\usepackage[english]{babel}
\usepackage{bm}
\usepackage{amssymb}
\usepackage{color, soul}
\usepackage{amsmath}
\usepackage{amstext}
\usepackage{latexsym}
\usepackage{graphicx}
\usepackage{mathtools}

\usepackage[ruled]{algorithm2e}
\begin{document}
	\author{Habib Ullah}
	\thanks{These  authors contributed equally to this paper.}
	
	\affiliation{College of Physics, Taiyuan University of Technology, Shanxi 030024, China}
	\author{Kun Li}
	\thanks{These authors contributed equally to this paper.}

	\affiliation{College of Physics, Taiyuan University of Technology, Shanxi 030024, China}


	\author{Haoyu Lu}
	
	\affiliation{College of Physics, Taiyuan University of Technology, Shanxi 030024, China}
	
	\author{Youjin Deng}%
	\email{yjdeng@ustc.edu.cn}
	\affiliation{
		Hefei National Laboratory for Physical Sciences at the Microscale and Department of Modern Physics, University of Science and Technology of China, Hefei 230026, China}
	\affiliation{Hefei National Laboratory, University of Science and Technology of China, Hefei 230088, China}
	
	\author{Wanzhou Zhang} \email{zhangwanzhou@tyut.edu.cn}
	\affiliation{College of Physics, Taiyuan University of Technology, Shanxi 030024, China}

	\date{\today}
	\title{Spiral states, first-order transitions and specific heat multipeak phenomenon in $J_1$-$J_2$-$J_3$ Ising model: A Wang-Landau algorithm study}
	
	\begin{abstract}
		The classical $J_1$-$J_2$-$J_3$ Ising model on the honeycomb lattice is important for understanding frustrated magnetic phenomena in materials such as FePS$_3$ and Ba$_2$CoTeO$_6$, where diverse phases (e.g., striped, zigzag, armchair) and magnetization plateaus have been experimentally observed. To explain the experimental results, previous mean-field studies have explored its thermal phase transitions, identifying armchair phases and striped phases, but their limitations call for more reliable numerical investigations.
		In this work, we systematically revisit the classical \(J_1\)--\(J_2\)--\(J_3\) Ising model using the Wang--Landau algorithm; in strongly frustrated parameter regimes where the Wang--Landau sampling becomes difficult, we further employ parallel tempering and a two-replica cluster algorithm combined with population annealing. We find that the armchair (AC) phase, previously reported in mean-field and experimental studies, actually coexists with the spiral (SP) phase, with their combined degeneracy reaching 20-fold (4-fold for the AC states and 16-fold for the spiral states). The phase transitions and critical exponents are studied at different interaction values. We observe first-order phase transitions, continuous phase transitions, and even the multipeak phenomenon in frustrated systems.
		These results clarify the nature of phases and phase transitions in frustrated Ising systems and their exponents, and additionally provide inspiration for experimental efforts to search for the spiral state and the specific-heat multipeak phenomenon.
		
	\end{abstract}
	
	\maketitle
	\section{Introduction}
	
	
	Frustrated spin models have long been a central topic in condensed matter physics~\cite{frustrated_book}. A prototypical example is the $J_1$-$J_2$-$J_3$ spin model, which incorporates nearest-neighbor ($J_1$), next-nearest-neighbor ($J_2$), and third-nearest-neighbor ($J_3$) interactions. On honeycomb lattices, the quantum variant of this model—such as the Heisenberg model—exhibits a rich array of phenomena, including spiral spin states~\cite{spiral2011,j1j2j3-q-heisenbg-3,q-spiral1,q-spiral2}. In particular, even when $J_3=0$, the system sustains a spin liquid phase~\cite{spinliquid,spinliquid2}, while its spin-1 counterpart displays distinct magnetic behavior~\cite{j1j2-spin1}. The study of these quantum systems has been significantly advanced by numerical techniques, particularly the density matrix renormalization group~\cite{dmrg} and tensor network approaches~\cite{q-spiral2,tn2}.
	
	Without quantum fluctuations, the classical $J_1$-$J_2$-$J_3$ Ising model was used to describe magnetic materials, such as FePS$_3$, a prime example of a two-dimensional Ising-like antiferromagnet on a honeycomb lattice formed by Fe$^{2+}$ ions \cite{wildes2020}. Other experiments on the honeycomb lattice, such as those on Ba$_2$CoTeO$_6$~\cite{experiment2}, where Co$^{2+}$ ions form the honeycomb lattice, have also been reported.  A variety of spin-density wave phases have been observed in these experimental systems, such as the striped phase, the zigzag phase, the armchair (AC)  phase, and various magnetization plateaus upon the application of a magnetic field \cite{wildes2020}. 
	
	To explore these phenomena, pioneering theoretical studies using the mean-field (MF) approach in Refs.~\cite{th1, zigzag} have examined the thermal phase transitions of the $J_1$-$J_2$-$J_3$ Ising model in the honeycomb lattices.
	In the range $J_3/J_1<1/2$,  the armchair phase is found. In the range $J_3/J_1>1/2$, rich critical and tricritical behavior is found. Due to the limitations of mean-field methods, the authors of Ref.~\cite{th1} also call on researchers to use more powerful Wang-Landau (WL) algorithms or parallel tempering (PT) algorithms~\cite{PhysRevLett.86.2050,fugao2,mpi-wanglandau,Shchur,tongninghua} to further explore the nature of phase transitions in this system~\cite{logarithic-windows}. In fact, exploring such frustrated systems on various lattices poses enormous challenges due to their strong ground-state degeneracy~\cite{Colbois_2022}. 
	\begin{figure}[t]
		\includegraphics[width=\linewidth]{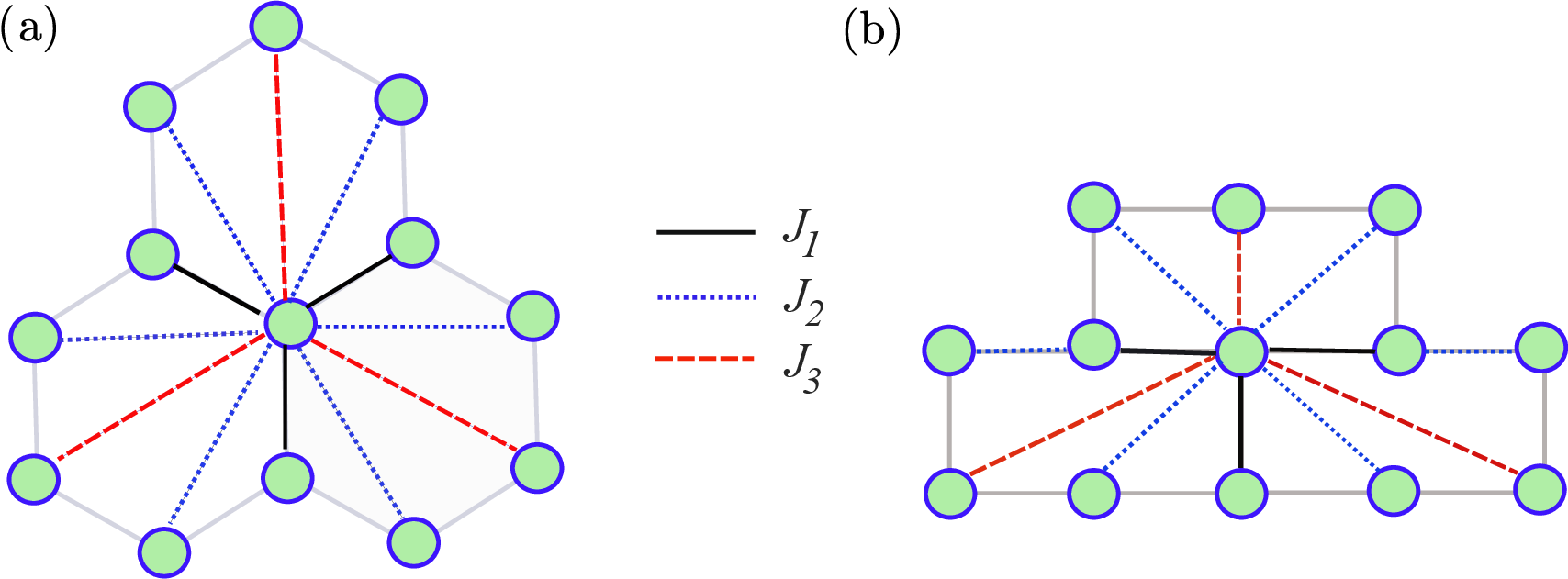}
		\caption{The lattice structure of a honeycomb lattice is shown in (a) for the hexagonal representation and in (b) for the brick-wall lattice, with \( J_1 \), \( J_2 \), and \( J_3 \) representing the nearest-neighbor, second-nearest neighbor, and third-nearest neighbor interactions, respectively.
		} 
		\label{Fig:lattice}
	\end{figure}

	In this paper, we systematically revisit  the $J_1$-$J_2$-$J_3$ model on the honeycomb lattice using the WL algorithm~\cite{PhysRevLett.86.2050,fugao2,mpi-wanglandau,Shchur,tongninghua}.
	In the strong frustration region, the PT method is also used as an auxiliary approach.

	The main findings are as follows. The AC phase revealed in MF studies~\cite{th1, zigzag} and experiments~\cite{wildes2020} coexists with the spiral (SP) phase in the intermediate parameter regime: the AC state is 4-fold degenerate, while the spiral state is 16-fold degenerate, leading to a total 20-fold degeneracy.
	Another key result is that, guided by the MF phase diagram, we accurately and reliably determine the ST-PM and AC+SP-PM phase transition boundaries at $J_3/J_1 = 1.2, 0.6, 0.2$, identify their transition types, and derive the corresponding critical exponents via the Wang-Landau algorithm. 
	Additionally, in the strong frustration regime $J_2/J_1=-1.5$ and $J_3/J_1=0.2$, the specific heat exhibits multi-peak behavior~\cite{multi_peaks}.

	The outline of our paper is as follows.
	Sec.~\ref{sec:model} presents the $J_1$-$J_2$-$J_3$ model, the WL method, and the measured quantities.
	The results on the phase diagram and details are shown in Sec.~\ref{sec:res}. The conclusion and discussion are given in Sec.~\ref{sec:con}.

	\section{The model, method and quantities}
	\label{sec:model}
	
	\subsection{$J_1$-$J_2$-$J_3$ Ising model}
	The Hamiltonian is given below.
	\begin{align}
		H = -\sum_a J_a\sum_{\langle i,j \rangle_a} S_i S_j,
	\end{align}
	where $a=1,2,3$ denotes the nearest-neighbor, next-nearest-neighbor, and third-nearest-neighbor interactions, as shown in Fig.~\ref{Fig:lattice}.
	From a programming perspective, compressing a hexagonal lattice into a brick-wall lattice is more straightforward to implement, as the coordinates and indices of the lattice sites are integers. Thus, we compress it into a brick wall lattice to discuss various phases.

	\subsection{Wang-Landau algorithm}
	In actual simulations, the cluster algorithm fails to capture the true correlations between spins and thus becomes ineffective~\cite{PhysRevLett.58.86}.  We also do not resort to the Metropolis algorithm because, at low temperatures, the system has a high degree of degeneracy, making it difficult to achieve ergodicity~\cite{Metropolis1953EquationOS}.
	
	Before the development of the WL method, several extended ensemble methods were proposed to improve sampling efficiency in systems with rough energy landscapes. These include the multicanonical ensemble~\cite{multicanonicalensemble, spin-glass}, entropic sampling~\cite{Entropic-sampling}, and the broad histogram Monte Carlo method~\cite{Broad-histogram}.
	The WL algorithm performs a random walk in the energy space by randomly flipping spin states~\cite{PhysRevLett.86.2050,fugao2,mpi-wanglandau,Shchur,tongninghua}. The energy \( E \) associated with the spin configuration is accepted with a probability proportional to the reciprocal of its density of states \( g(E) \), and \( g(E) \) is updated and recorded during the simulation. In the random walk, to traverse the entire energy space, a cumulative histogram \( H(E) \) is introduced: each time the energy \( E \) is visited, the corresponding value of \( H(E) \) increases by 1. Eventually, the visit distribution $H(E)$  for each energy \( E \) is obtained. A flat energy histogram indicates that each energy is visited uniformly.
	The WL algorithm has been continuously  evolving. People have been improving its accuracy, convergence speed, and so on from multiple perspectives~\cite{zhou2005understanding,brown2011convergence,caparica2012wang,morozov2007accuracy,poulain2006performances,zhou2006wang,berg2007wang,landau2004new,wang2001determining,perera2014replica}, and this algorithm has also been widely applied to frustrated spin or tricritical spin systems~\cite{silva2006wang,takeuchi2017new,qin2009two,shevchenko2017entropy,azhari2020tricritical,kwak2015first,Azhari_2022} and other aspects~\cite{liuwei1,liuwei2}. This method was also applied to the $J_1-J_2$ Ising model in generalized triangular lattices~\cite{Jang_2024}.
	
	The specific steps of the WL algorithm are as follows:  
	\begin{itemize}
		\item[(1)] Set the initial spin state of the entire lattice, initialize the density of states \( g(E) = 1 \) for all possible energies \( E \), and set a modification factor, for example, \( f = 2 \).  
		\item[(2)] Randomly flip a spin to form a trial state. If the energy changes from \( E_{\text{old}} \) (before flipping) to \( E_{\text{new}} \) (after flipping), the transition probability from \( E_{\text{old}} \) to \( E_{\text{new}} \) is given by:  
		\[
		P(E_{\text{old}} \to E_{\text{new}}) = \min\left( \frac{g(E_{\text{old}})}{g(E_{\text{new}})}, 1 \right) \tag{1}
		\]  
		If \( g(E_{\text{new}}) \leq g(E_{\text{old}}) \), the flipped state is accepted. Otherwise, it is randomly accepted with a probability of \( \frac{g(E_{\text{old}})}{g(E_{\text{new}})} \).  
		\begin{itemize}
			\item If the flipped state is accepted, the density of states is updated as \( g(E_{\text{new}}) \to g(E_{\text{new}}) \times f \), and the energy histogram is updated as \( H(E_{\text{new}}) \to H(E_{\text{new}}) + 1 \).  
			\item If the random walk state is rejected and the system remains at the same energy \( E_{\text{old}} \), the existing density of states is multiplied by \( f \), that is, \( g(E_{\text{old}}) \to g(E_{\text{old}}) \times f \), and the energy histogram is updated as \( H(E_{\text{old}}) \to H(E_{\text{old}}) + 1 \).  
		\end{itemize}
		Since \( g(E) \) can become very large, the algorithm uses the logarithm of the density of states in practice, i.e. \( \ln[g(E)] \to \ln[g(E)] + \ln(f) \).  
		\item[(3)] Continue to execute step (2) until the histograms flatten, then reduce the modification factor using \( f_1 = \sqrt{f} \).  
		\item[(4)] Repeat steps (2) and (3) until \( f = f_{\text{min}} \), and the final density of states \( g(E) \) is obtained.  
	\end{itemize}
	
	In practical simulations, there are many hyperparameters to choose from, such as the value of \(f_{\text{min}}\), the criteria for judging flatness, how many evolution steps to perform before each flatness check, and whether all energy levels are subject to flatness checks, etc. 
	
	\subsection{The parallel tempering method}
	
	PT is a simulation approach that boosts the efficiency of Monte Carlo (MC) sampling techniques~\cite{pt}, and it has proven particularly valuable in frustrated and spin-glass systems~\cite{binder_spinglass,pt_glass}.  
	We employ this method as it produces more dependable simulation outcomes when $J_3 = 0.2$.  
	The fundamental concept underlying the PT approach involves simulating multiple system replicas at distinct temperatures concurrently. Each replica serves as a copy of the system, featuring an identical set of spins but operating at a different temperature. The acceptance probability for an exchange between adjacent replicas adheres to:  
	\begin{equation}
		\begin{split}
			p       &=\min \left(1, e^{\left(E_{i}-E_{i+1}\right)\left(\frac{1}{ T_{i}}-\frac{1}{T_{i+1}}\right)}\right)
		\end{split},
		\label{eq:pb}
	\end{equation}  
	where $T_{i(i+1)}$ denotes the temperature of the respective replica, and $E_{i(i+1)}$ stands for the energy of the replica at temperature $T_{i(i+1)}$.
	The method has also been used to simulate the Ising-XY model successfully~\cite{Ma__2024}.

	\subsection{Population annealing algorithm with two-replica cluster updates}
	\label{sec:PA}
	
	The simulation~\cite{keisermachta2026} starts with parameter setup, lattice construction and topology validation, followed by initialization of $R$ random spin replicas and generation of a cooling schedule for inverse temperature \(\beta\).
	For each annealing step from \(\beta_\mathrm{old}\) to \(\beta_\mathrm{new}\):
	\begin{itemize}
		\item[1.] Compute the energy of all replicas.
		\item[2.] Perform population annealing resampling: replicas are duplicated or pruned according to Boltzmann weights, and an array is updated to track ancestry.
		\item[3.] Execute $S$ sweeps of two-replica cluster updates: replicas are randomly paired, cluster construction is implemented via breadth-first search, and spins within the cluster are flipped simultaneously for each paired replica pair.
		\item[4.] Carry out \(n_\mathrm{metropolis}\) sweeps of single-spin Metropolis updates.
		\item[5.] Evaluate thermodynamic observables including $E$, $e$, and \(C_V\), and record timing statistics.
	\end{itemize}
	After completing all \(\beta\) steps, the temperature-dependent data \((T,e,C_V)\) are written to text files and global simulation statistics are summarized.

	\subsection{The measured quantities}
		For the  WL algorithm, the measured quantities include the following:
		\begin{itemize}
			\item The energy \( U(T) \), which can be obtained using the relation:
			\begin{align}
				U(T) = \frac{\sum_E E \, g(E) \, e^{-E /  T}}{\sum_E g(E) \, e^{-E /  T}}
			\end{align}
			\item The specific heat \( C_V(T) \), derived from the internal energy through the thermodynamic relation:
			\begin{align}
				C_V = \frac{\text{d}U(T)}{\text{d}T} = \frac{\langle E^2 \rangle - \langle E \rangle^2}{ NT^2}
				\label{eq:cv}
			\end{align}
		\end{itemize}

		\section{Results}
		\label{sec:res}
		We first discuss the zero temperature phase diagram, as well as the configurations of each phase and the corresponding structure factors. Then, we proceed to discuss the finite-temperature phase diagram for $J_3/J_1=1.2$, 0.6, and 0.2, the exponents $y_t$ of continuous phase transitions, and evidence for first-order phase transitions and the specific heat multi-peak phenomenon.
		
		The zero-temperature phase diagram has been shown in Refs.~\cite{th1, zigzag} in the range $-2\le J_2/J_1\le  0$ and
		$0\le J_3/J_1\le 2$ and in a study of a real material~\cite{wildes2020}.
		To compare with previous works~\cite{th1, zigzag,wildes2020}, this paper also provides the zero-temperature phase diagram for this parameter range.
		
		\subsection{The results at $T=0$}
		\label{sec:T0}
		\subsubsection{
			The global phase diagram at $T=0$}
		
		\begin{figure}[t]
			\centering  \includegraphics[width=0.9\linewidth]{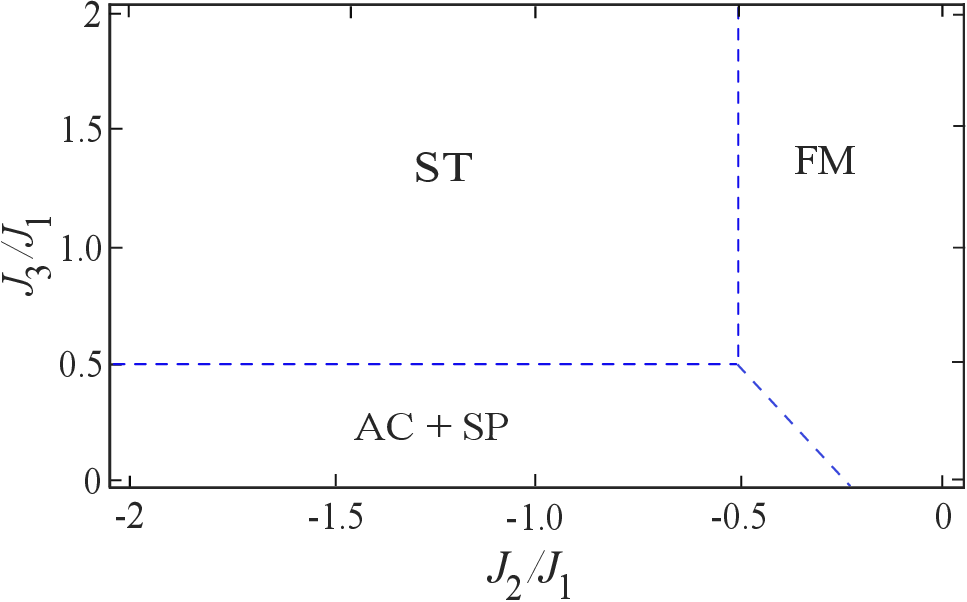}
			\caption
			{Phase diagram at zero temperature, which contains the ST, FM, and AC+SP phases. Different from Ref.~\cite{th1}, we find the AC phase should coexist with the SP phase.}
			\label{fig:T0}
		\end{figure}
		
		\begin{figure}[t]
			\centering  \includegraphics[width=\linewidth]{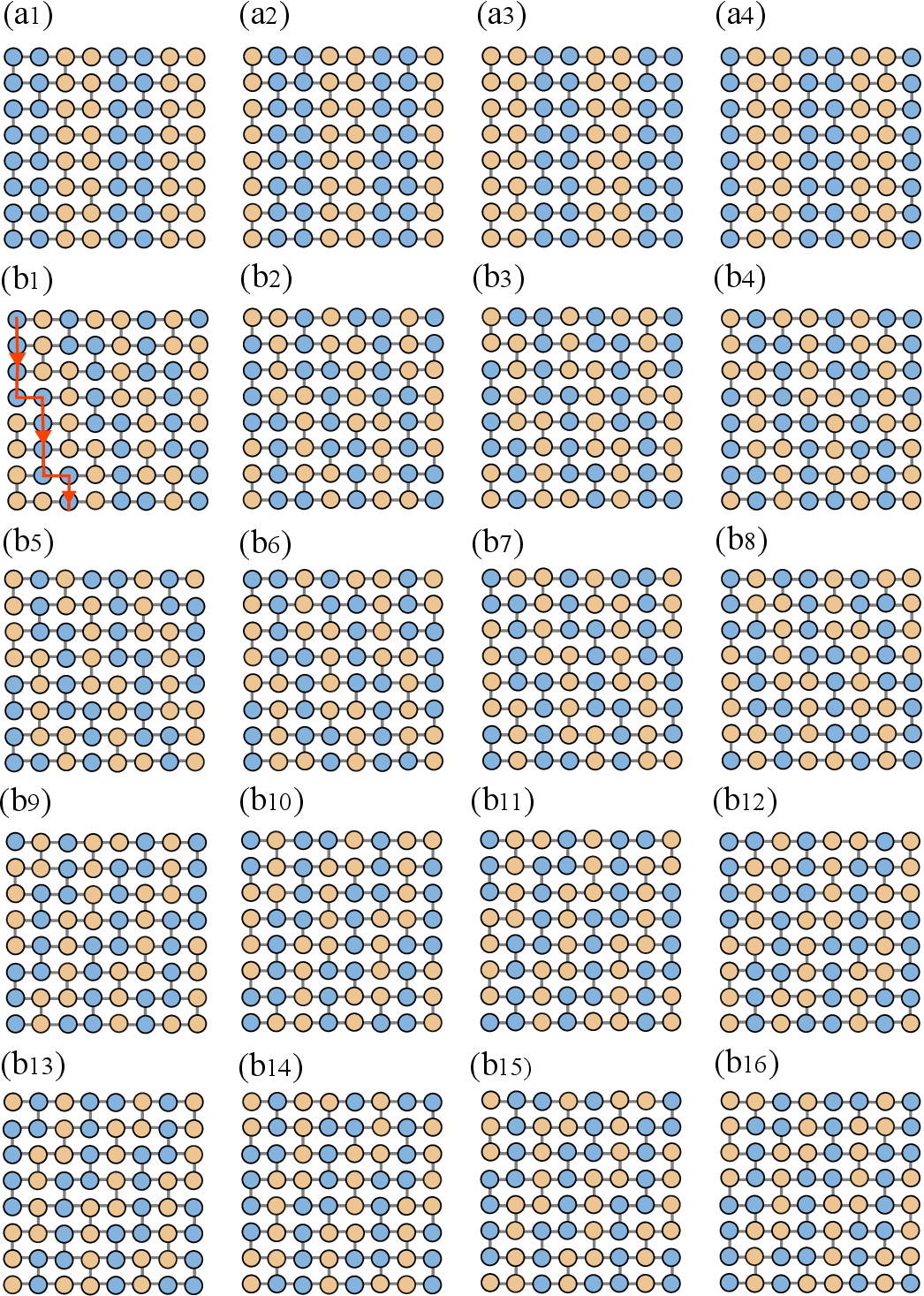}
			\caption
			{Ground state spin configurations obtained from WL sampling. (a1)-(a4) show  configurations of the AC phase, while  (b1)-(b16) correspond to various spiral configurations. These 20 configurations are degenerate, which has not been found in previous works~\cite{wildes2020,th1, zigzag}. Red lines and arrows in (b1) indicate the spiral direction of spin configurations.
			}
			\label{fig:ac_confg}
		\end{figure}

			Figure~\ref{fig:T0} presents the phase diagram of the  ground state, which includes the stripe phase,  the AC+SP phase, and the FM phase. The boundaries of the three phases can be determined analytically from energy calculations. The energy expressions per site for the three phases are as follows:
			\begin{align}
				E_{\text{AC+SP}} &= -\frac{J_1 - 2J_2 - J_3}{2}, \label{eq:energy_ac} \\
				E_{\text{ST}} &= \frac{J_1 + 2J_2 - 3J_3}{2}, \label{eq:energy_st} \\
				E_{\text{FM}} &= -\frac{3J_1 + 6J_2 + 3J_3}{2}. \label{eq:energy_fe}
			\end{align}

		Setting \(E_{\text{AC+SP}} = E_{\text{ST}}\), we obtain the boundary between the AC and ST phases as \(J_3/J_1 = 1/2\). By setting \(E_{\text{AC+SP}} = E_{\text{FM}}\),  the boundary between the AC+SP and FM phases is given by \(J_3/J_1 = -2 J_2/J_1 - 1/2\). Furthermore,  the ST-FM boundary is given by \(J_2/J_1 = -1/2\) using the same method.
		
		\subsubsection{The typical configurations}
		
		In Fig.~\ref{fig:ac_confg}, the ground-state spin configurations obtained from the WL sampling are shown. Figures~\ref{fig:ac_confg} (a1)-(a4) show representative configurations of the AC phase, while Figs.~\ref{fig:ac_confg} (b1)-(b16) correspond to various configurations in the SP phase. Orange and blue dots indicate spin-up and spin-down orientations, respectively.
		It can be clearly seen that the degeneracy of the AC phase is four. Specifically, configurations (a2)-(a4) can be obtained by sequentially shifting the stripes of (a1) to the right, which reflects the invariance under horizontal translation. Due to the asymmetry between the horizontal and vertical directions of the lattice structure, these four vertical stripes do not have horizontal counterparts, unlike those of the square lattice~\cite{j1j2sq}. 
		In  Figs~\ref{fig:ac_confg} (a1)-(a4), the  Hamiltonian exhibits space discrete translational symmetry along the horizontal-direction, satisfying $\mathcal{T}_x^1 H \mathcal{T}_x^{-1}=H$, and also possesses $\mathbb{Z}_2$ symmetry under global spin flip.
		
		\begin{figure}[t]
			\centering  \includegraphics[width=\linewidth]{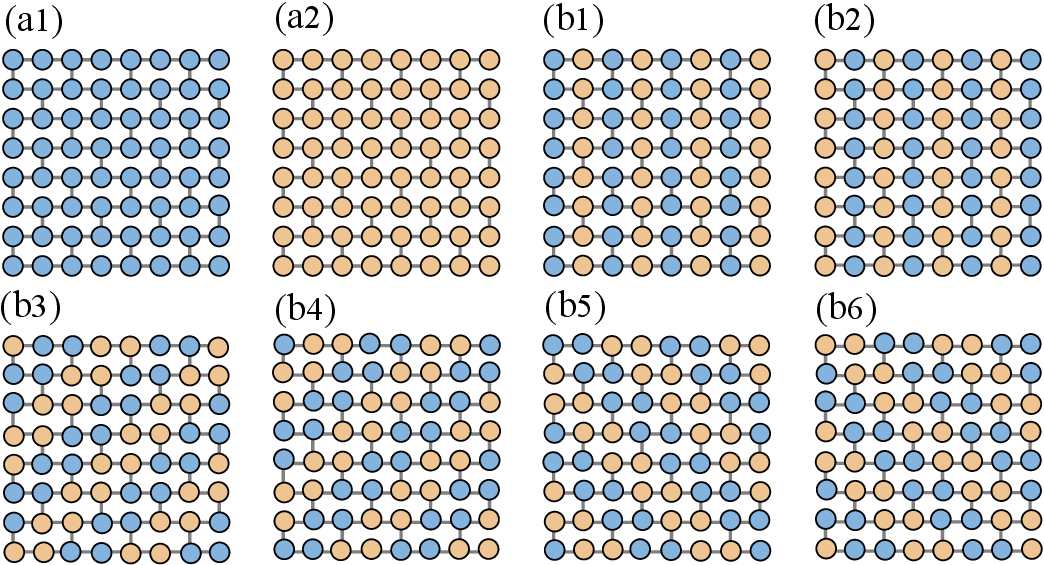}
			\caption
			{Ground-state spin configurations for (a1)-(a2) the FM phase and (b1)-(b6) the ST phase.}
			\label{fig:st_confg}
		\end{figure}

			In Figs.~\ref{fig:ac_confg} (b1)-(b16), the spiral configurations are shown. 
			The red lines and arrows indicate the spiral direction of spin configurations.
			In Fig.~\ref{fig:ac_confg} (b1), the first four elements in the first column are blue circles. Then, the fourth blue circle moves to the second column, followed by another four blue circles displayed downward. At the end of the blue column, it moves to the right again. Due to the periodic boundary condition, there are only 2 blue circles at the bottom of the third column, while the other two blue circles roll to the topmost position. In this way, the blue circles spread throughout the lattice and eventually return to the starting point. Figs.~\ref{fig:ac_confg} (b2)-(b8) are similar by moving the first four blue circles down one by one. 
		In Figs.~\ref{fig:ac_confg} (b9)-(b16), the configurations can be obtained by applying the $x$-coordinate mapping \(x \to 9-x\) to the configurations shown in Figs.~\ref{fig:ac_confg}(b1)-(b8).
		In this Appendix~\ref{sec:app}, a specific SP
		configuration on the hexagonal lattice  is shown.

		From a symmetry perspective, the configurations exhibit discrete translational symmetry along the vertical direction. For any single-step vertical translation operation $\mathcal{T}_y$, the energy of the configuration is invariant under its action, satisfying 
		\begin{equation}
			\mathcal{T}_y^n \mathcal{C} \mathcal{T}_y^{-n} = \mathcal{C},\quad E(\mathcal{C}) = E(\mathcal{T}_y^n \mathcal{C}),
		\end{equation}
		Here  $\mathcal{C}$ represents any configuration in the first subset, $n = 1,2,\dots,8$ is the number of translation steps, and $E(\mathcal{C})$ denotes the energy of configuration $\mathcal{C}$.
		The configurations also possess reflection symmetry with
		\begin{equation}
			\mathcal{R}_{x \mapsto 9-x} \mathcal{C}_i \mathcal{R}_{x \mapsto 9-x}^{-1} = \mathcal{C}_i^\prime. 
		\end{equation}
		Here $\mathcal{R}_{x \mapsto 9-x}$ is the mirror reflection operator, with $\mathcal{C}_i$ ($i=1,\dots,8$) corresponding to $b_1$–$b_8$ and $\mathcal{C}_i^\prime$ ($i=1,\dots,8$) denoting Figs.~\ref{fig:ac_confg} $(b_9)$–$(b_{16})$.
		
		The spiral state has the properties defined in
		Refs.~\cite{spiral2011,j1j2j3-q-heisenbg-3,q-spiral1,q-spiral2}. Generally, the spin structure factors of these spiral spin states exhibit discrete bright spots in momentum space at low temperatures, while forming rings at high temperatures or under strong quantum fluctuations.
		The contour of the ring in momentum space can be observed experimentally~\cite{ex1, ex2}.
		The spiral state in this paper is defined differently. As defined in Refs.~\cite{pmq1pmq2, q-spiral2},
		at lower temperature, the peaks of the spin structure factor in momentum space are located at $(\pm k_1, \pm k_2)$, and in the real space, the configuration has spiral properties.

			In Fig.~\ref{fig:st_confg}, we also present the configuration of the FM phase and the ST phase. Among the ST phase, the stripes in Figs.~\ref{fig:st_confg} (b1) and (b2) are along the vertical direction, with the spin directions of the odd columns and even columns being opposite to each other. Figs.~\ref{fig:st_confg}(b3) and (b4) also show the ST phase, where the stripe width is 2 and the stripes are tilted at a 45-degree angle, similar to the shape of ``//". In Figs.~\ref{fig:st_confg} (b5) and (b6),  the tilt direction is like ``$\backslash\backslash$".
			In (b3)-(b6), these four configurations possess not only $\mathbb{Z}_2$ symmetry but also $90^\circ$ rotational symmetry.

		\subsubsection{The spin structural factors}

		\begin{figure}[t]
			\centering  \includegraphics[width=1\linewidth]{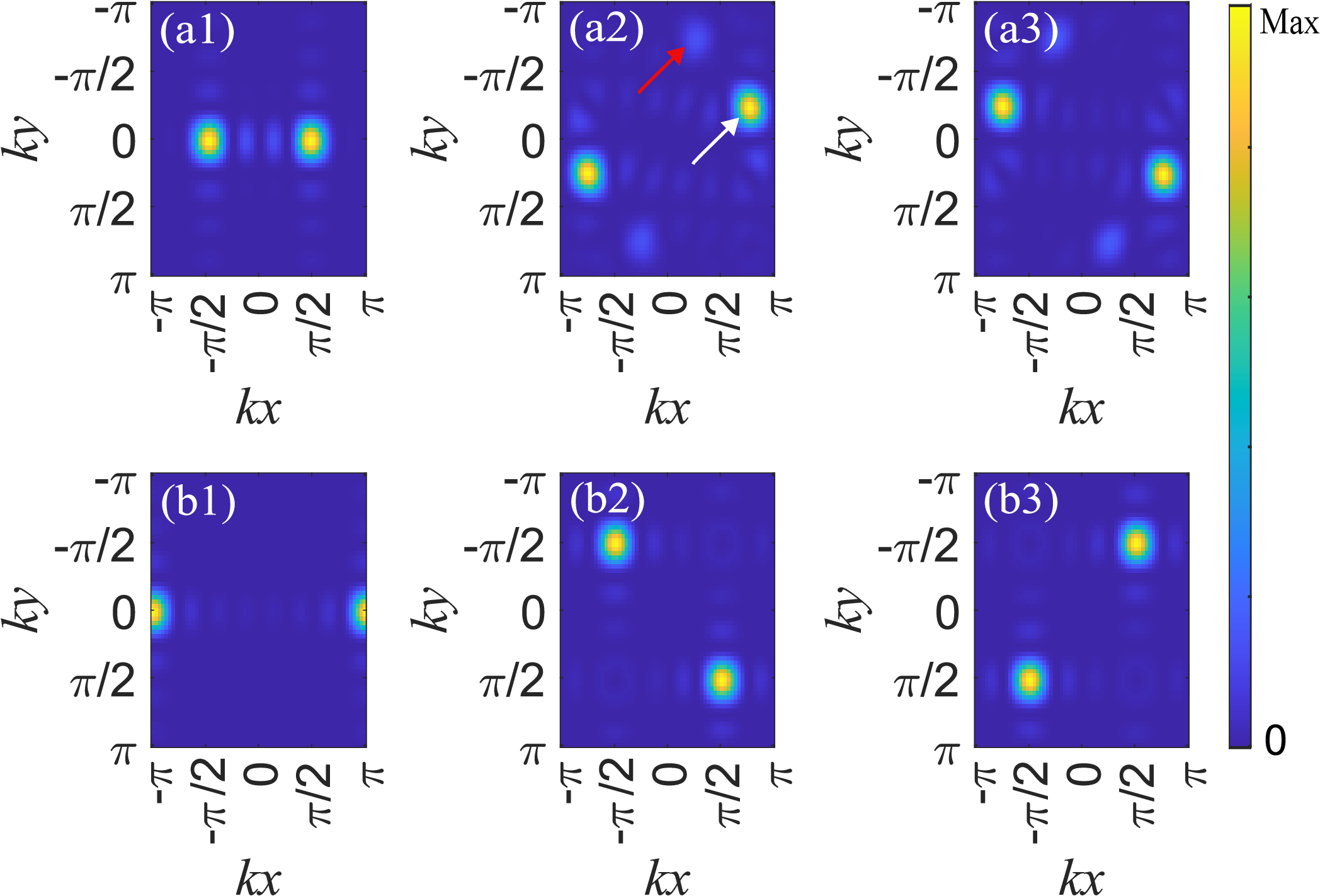}
			\caption
			{Structure factors corresponding to different configurations: (a1) for the AC phase; (a2) and (a3) are for the SP configurations; (b1)-(b3) for the ST configurations.}
			\label{fig:sq}
		\end{figure}

		
		To further characterize the phases, we define the spin structure factor for the spins as
		\begin{equation}S(k_x,k_y)=\frac{1}{N} \sum_{i,j}^{N} e^{i\bm{k}\cdot(\bm{r}_i-\bm{r}_j)}\left\langle S_iS_j \right\rangle,
			\label{eq:sq}\end{equation}
		where $N$ is the number of spins, $\bm{r}_i$ and $\bm{r}_j$ are the coordinates of the spins, and $S_i$ and $S_j$ are the spin values.  
		In the analysis of magnetic phases, the structure factor $S(\vec{\bm{k}})$ provides critical insight into the spatial ordering of spin configurations. The presence of peaks in reciprocal space is directly related to the periodicity and symmetry of the real-space spin arrangements.

		In Fig.~\ref{fig:sq}, structure factors corresponding to different configurations are shown.
		In Fig.~\ref{fig:sq} (a1), the bright spots at $(\pm \pi/2, 0)$ correspond to the four AC configurations depicted in Figs.~\ref{fig:ac_confg} (a1)-(a4). In the AC state, the translation-invariant period length is $\Delta x_{\text{AC}} = 4$, and the relation $k = \pm 2\pi/\Delta x$ results in the wave-vector positions $k_x=\pm \pi/2$.
		This solid phase with a period of 4 has also been observed in Rydberg atoms with long-range interactions~\cite{atom1,atom2}.
		In Fig.~\ref{fig:sq} (a2),  the bright spot indicated by the white arrow shows peaks at $k_y = \pm \pi/4$, and the bright spot indicated by the red arrow shows peaks at $k_x =\pm \pi/4$. These vectors $(\pm \pi/4, \pm \pi/4)$ correspond to the translation-invariant wave vector where $\Delta x= 8$ or $\Delta y = 8$ of the SP configurations, similar to the results  in Refs.~\cite{pmq1pmq2, q-spiral2}.
		In Fig.~\ref{fig:sq} (b1), bright spots appear at $(\pm \pi, 0)$, arising from the two vertical stripes in the ST phase, where $\Delta x_{\text{ST}}= 2$.
		In Figs.~\ref{fig:sq} (b2) and (b3), the bright spots at $(\pm \pi/2, \pm \pi/2)$ signify the four diagonal stripes shown in Figs.~\ref{fig:st_confg} (b3)-(b6).
		The data in the figure are the results for $L=8$,
		and other smaller dark spots are caused by finite-size effects and will disappear as the system size increases.
		
		\subsection{The phase diagram at $T>0$}
		\label{sec:T>0}
		\subsubsection{$J_3/J_1=1.2$}
		
		\begin{figure}[t]
			\centering  \includegraphics[width=\linewidth]{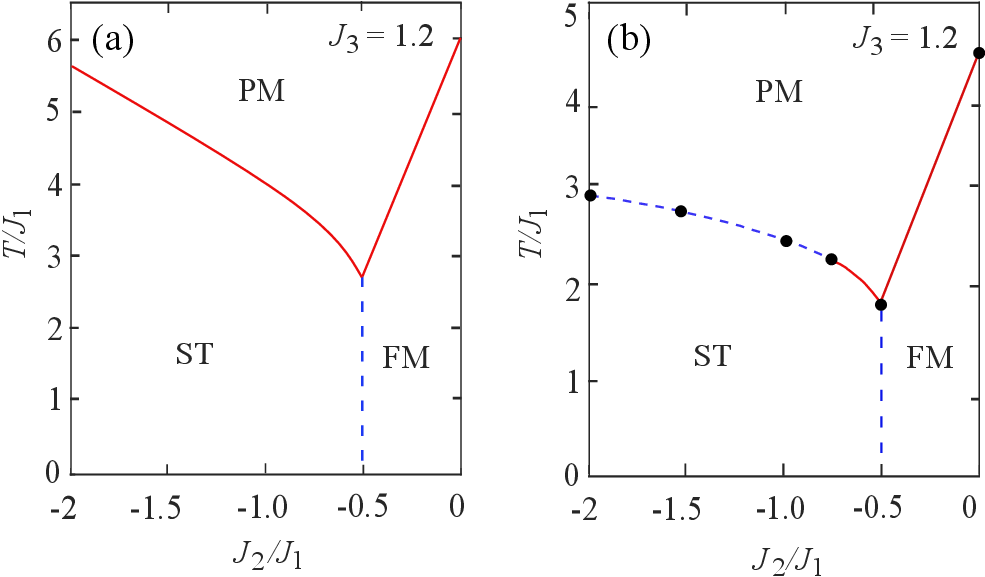}
			\caption
			{Finite-temperature phase diagram of $J_3/J_1=1.2$ (a) from Ref.~\cite{th1} by the MF method (b) by the WL method. Solid lines indicate continuous phase transitions, whereas dashed lines represent first-order transitions. The tricritical point is about $J_2^c/J_1\approx -0.75$.}
			\label{fig:phase_J3_1.2}
		\end{figure}
		
		\begin{figure*}[htbp!]
			\includegraphics[width=\linewidth] {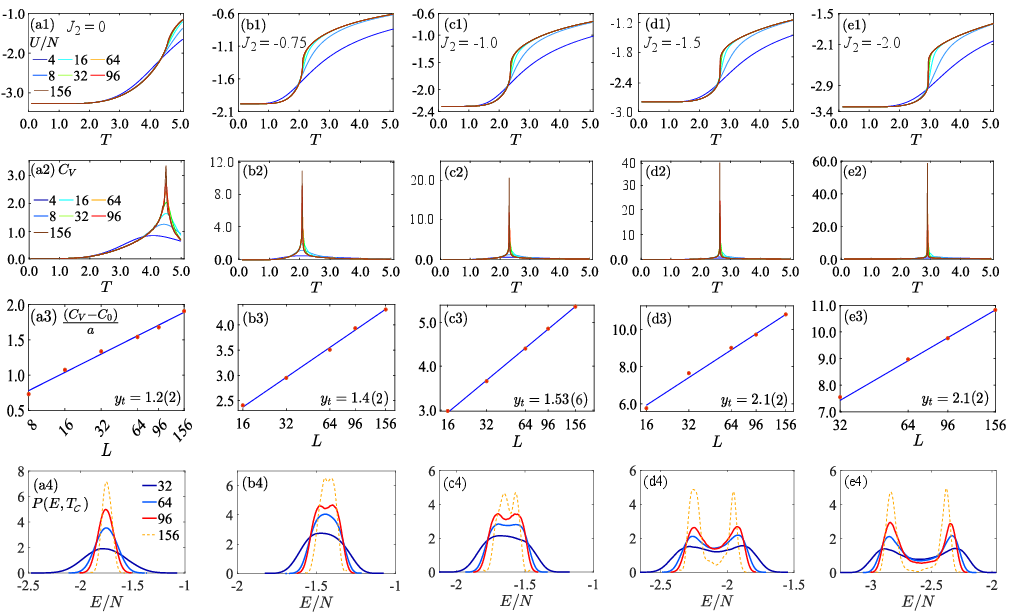}
			\caption{{Details $U/N$, $C_V$, fitting of exponent $y_t$, and normalized distribution $P(E/N, T_c)$ are shown for  $J_3=1.2$. For each column, the parameters are $J_2/J_1 = 0, -0.75, -1, -1.5, -2$, respectively. The columns (a) and (b) correspond to continuous phase transitions; column (c) corresponds to a weak first-order transition; while the last two columns correspond to first-order phase transitions.}}
			\label{fig:J3=1.2}
		\end{figure*}

			Based on the zero-temperature phase diagram in the \( J_3/J_1 \)-\( J_2/J_1 \) parameter space, we fix \( J_3 = 1.2 \) and plot the finite-temperature phase diagram in the \( J_2/J_1 \)-\( T/J_1 \) parameter space.
			For ease of comparison, we have included the phase boundaries of the MF method in Refs.~\cite{th1, zigzag} in Fig.~\ref{fig:phase_J3_1.2}(a), while the phase boundaries obtained from the WL algorithm are shown in Fig.~\ref{fig:phase_J3_1.2}(b). Solid lines represent continuous phase transitions, and dashed lines represent first-order phase transitions.
			In general, the topological structure of the phase diagrams is the same. The ST phase is located in the lower-left corner, the FM phase in the lower-right corner, and the paramagnetic phase in the high-temperature region.
			
			There are  differences in the nature of the phase transitions between the two methods. For example, in the region where \(J_2/J_1 < J_2^c/J_1\) and $J_2^c/J_1\approx -0.75$, the WL method reveals that the system undergoes a  first-order phase transition,  rather than the continuous phase transition identified by the MF algorithm.
			The strength of the WL algorithm lies in its ability to calculate the density of states more accurately,  allowing the calculation of energy and its distribution $P(E,T)$ at different temperatures $T$. The double peaks at $T=T_c$ confirm the first-order transitions. 
			For the range $J_2/J_1 > J_2^c/J_1$, the transitions are confirmed  {\it continuous}. 
			Sharp peaks in the specific heat help us to precisely obtain the critical exponent \( y_t \).
			
			Calculations of the critical exponents reveal that the phase transition type does not necessarily belong to the Ising universality class, since the ground state of the system has a degeneracy of six (the ST phase), as shown in Fig.~\ref{fig:st_confg}.
			
			For a first-order phase transition, a bimodal distribution can be obtained by finely tuning the temperature $T$ and selecting an appropriate critical temperature \(T=T_c\).
			As shown in Figs.~\ref{fig:J3=1.2} (b4) and (c4), for the largest system size achievable  \(L = 156\), a weak bimodal energy distribution is observed.

			In Fig.~\ref{fig:J3=1.2}, details of fixing $J_3=1.2$ and scanning $T/J_1$ are shown. 
			{The first row shows the relationship between the average energy $U/N$ and temperature for $J_2/J_1 = 0, -0.75, -1, -1.5$, and $-2$.}
			At low temperatures ($T<T_c$) and high temperatures (e.g., $T=5$), the energies of different sizes ($L\ge8$) are convergent, as expected.
			Meanwhile, the values of these ground-state energy densities must be in accordance with Eq.~\ref{eq:energy_st}.
			Each curve of energy versus temperature has undergone strict convergence testing, such as increasing the number of iteration steps, imposing stricter flatness conditions for histogram $H(E)$, and using values of $f_{\text{min}}$ closer to 1.

			In Fig.~\ref{fig:J3=1.2}, the second row presents the relationship between specific heat $C_V$ and temperature $T$.
			We divide the temperature range from 0 to 5  with an interval of $5\times 10^{-5}$. Then, according to Eq.~\ref{eq:cv}, we calculate the specific heat for different sizes. Each specific heat curve has a peak. The critical temperature \(T_c(L)\), determined from the peaks of the specific heat, is retained in the fourth decimal place; Then, using the software ``gnuplot'', the phase transition point in the thermodynamic limit is obtained via linear fitting with the least-squares method, recorded in Table~\ref{tab:yt1.2}, and serves as a reference for other methods, such as tensor network methods~\cite{j1j2trg,j1j2trg2,tnmc2,chen2025,chen2025twodimensional,tnmc}.
			
			
			In  Fig.~\ref{fig:J3=1.2}, the third row, data points $(C_V^{\text{max}}(L) - C_0)/a$ versus $L$ are plotted on a log-log scale, with those of different sizes all lying on the fitted straight line.
			The slope of this fitted straight line is governed by the following finite-size scaling form~\cite{cvfit}, from which the critical exponent \( y_t \) is extracted:
			\begin{equation}
				C_V^{\text{max}}(L) = C_0 + L^{2y_t - d}(a + bL^{y_1}), 
				\label{eq:Cv}
			\end{equation}
			where \( C_V^{\text{max}}(L) \) is the specific-heat maximum to be fitted, \( L \) is the size of the lattice, \( C_0 \), \( a \), and \( b \) are constants, and \( d \) is the spatial dimension. In the fitting process, $b=0$ is used, and the specific heat values used are the maximum values of the specific heat $C_V^{\text{max}}$ for different lattice sizes.
			The fitted results of $y_t$ are shown in Table~\ref{tab:yt1.2}. The actual phase-transition type should be judged comprehensively by further checking whether a stable double-peak (bimodal) exists in the energy distribution shown in the fourth row of Fig.~\ref{fig:J3=1.2}. 
			
			Below, we apply this criterion to each $J_2/J_1$ value and discuss their phase-transition types in turn:
			\begin{itemize}
				\item $J_2/J_1=0$: The system exhibits a standard Ising FM–PM transition upon heating. From Eq.~\ref{eq:Cv}, we get $y_t=1.2(2)$, consistent with expectation. The Ising relation $\alpha = 2y_t-d=0$ gives logarithmic divergence of $C_V^{\text{max}}$~\cite{ising},
				\begin{equation}
					C_V^{\text{max}}(L)=C_0+C_1\ln L,
				\end{equation}
				which we use to fit the peak specific-heat data. The energy distribution is unimodal, as shown in Fig.~\ref{fig:J3=1.2}(a4).
				
				
				\item \(J_2/J_1=-0.75\): {The fitted critical exponent is \(y_t=1.4(2)\), smaller than the first-order value \(2\). For a weakly first-order transition, however, \(y_t\) converges to \(2\) only in the large-system-size limit~\cite{j1j2sq}; the present value alone therefore does not allow a definite conclusion. Although we have reached \(L=156\), the system size may still be insufficient, to settle whether this transition is weakly first order. Judging from the energy distribution, as shown in Fig.~\ref{fig:J3=1.2}(b4), an energy bimodal distribution is observed, but the width of the bimodal peaks gradually decreases with increasing system size. The transition thus remains continuous. This behavior is highly analogous to that of the four-state Potts model reported in Ref.~\cite{j1j2sq}, which is known to exhibit a continuous phase transition.}
				
				\item $J_2/J_1=-1.0$: {The fitted critical exponent is $y_t=1.53(6)$, smaller than 2. This points to a {weak first-order} transition for $L\le 156$. Consistent with this, a weak bimodal energy distribution is seen at our largest size $L=156$, as shown in Fig.~\ref{fig:J3=1.2}(c4).}

				\item $J_2/J_1=-1.5,-2.0$: Both yield $y_t=2.1(2)$, consistent with 2 within three standard deviations. Inserting $y_t$ into Eq.~\ref{eq:Cv}, the specific heat diverges as $L^d$, characteristic of a clear first-order transition~\cite{Sandvik_2010}. The prominent double peaks shown in Figs.~\ref{fig:J3=1.2}(d4) and (e4) confirm this behavior.
			\end{itemize}

			
			\begin{table}[tbh]
				\caption{Fitting results for the $C_V$ using the ansatz Eq.~\ref{eq:Cv} at $J_3/J_1=1.2$. The numbers in parentheses indicate errors, representing three times the standard deviation.
				}
				\begin{tabular}{l@{\hspace{14pt}}l@{\hspace{14pt}}l@{\hspace{14pt}}l}
					\hline\hline
					${J}_{2}/{J}_{1}$ ~~~~~& ~~${y}_{t}$ ~~~~& ~~~~${T}_{c}$~~~~~~  &Type\\ \hline
					0 & 1.2(2) & 4.513(9) & continuous \\
					-0.75 & 1.4(2) & 2.085(3) &continuous\\
					-1.0 & 1.53(6) & 2.316(2) &weak first order\\
					-1.5 & 2.1(2) & 2.643(2) & first order\\
					-2.0 & 2.1(2) & 2.876(2) & first order\\ \hline
					\hline
					\label{tab:yt1.2}
				\end{tabular}
			\end{table}

			

			
			
			\subsubsection{$J_3/J_1=0.6$}
			
			\begin{figure}[t]
				\centering  \includegraphics[width=\linewidth]{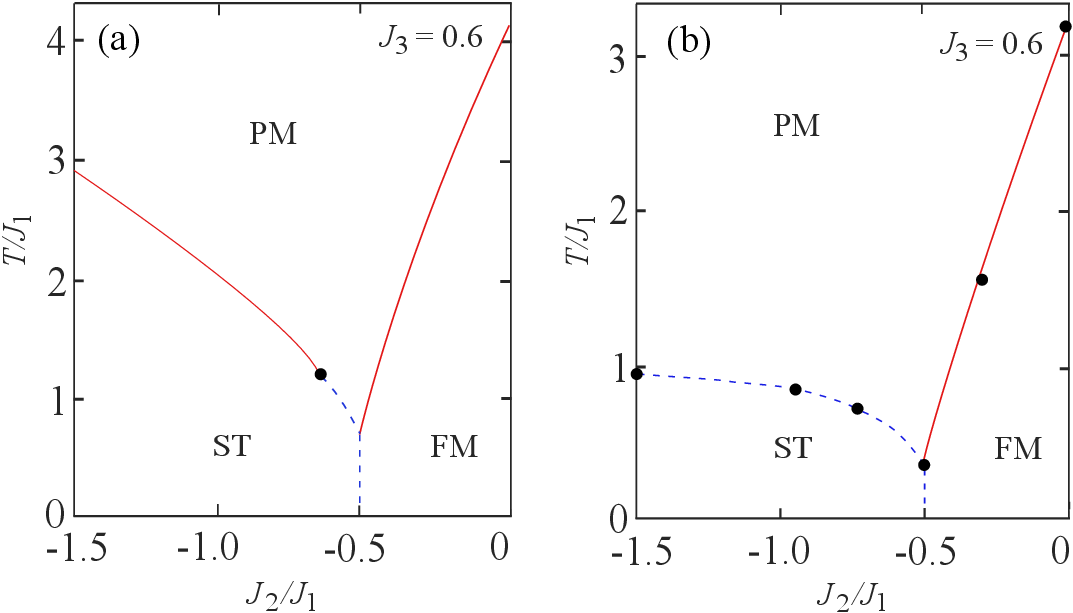}
				\caption
				{Finite temperature phase diagram of $J_3=0.6$ (a) from Ref.~\cite{th1} by the MF method (b) by the WL method. Solid lines indicate continuous phase transitions, whereas dashed lines represent first-order transitions.}
				\label{fig:phase_J3_0.6}
			\end{figure}

			\begin{figure}[hbt]
				\centering   \includegraphics[width=1\linewidth]{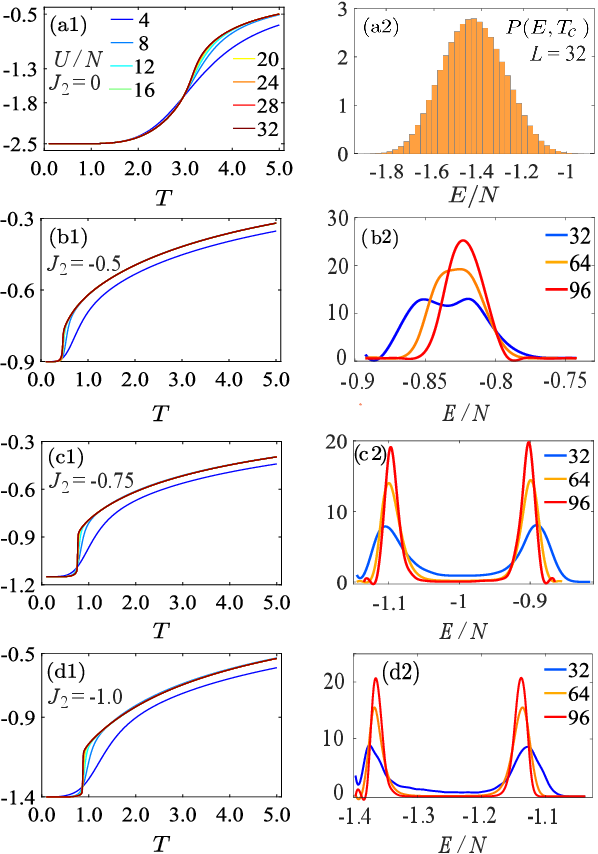}
				\caption
				{
					Rows correspond to $J_2/J_1=0, -0.5, -0.75$, and $-1.0$, respectively, with $J_3/J_1=0.6$ fixed.
					The first column shows $U/N$ versus $T$, and the second column shows the energy distribution at the phase transition point.
					For the top two rows the transition is continuous: at $J_2/J_1=0$ the distribution has a single peak, while at $J_2/J_1=-0.5$ the double peaks vanish as $L$ increases.
					For the bottom two rows ($J_2/J_1=-0.75$ and $-1.0$), the double peak persists for all system sizes and, together with the jump of $U/N$, indicates a first-order phase transition.}
				\label{fig:J3=0.6}
			\end{figure}

			We now fix \( J_3/J_1 = 0.6 \) and plot the finite-temperature phase diagram.  This parameter is closer to the boundary $J_3/J_1=0.5$ at $T=0$ in Fig.~\ref{fig:T0}, and therefore the transitions may be different from those at $J_3/J_1=1.2$ discussed in the previous section. Actually, for the ST-PM phase transition, the WL results differ significantly from the mean-field results: MF finds a tricritical point, whereas WL reveals that it is first-order phase transition.

			In Fig.~\ref{fig:phase_J3_0.6} (a),  the phase boundaries are obtained using the MF method, while Fig.~\ref{fig:phase_J3_0.6} (b) presents the results of the WL algorithm.  The ST, FM, and PM phases occupy consistent positions in the two phase diagrams.
			The nature of phase boundaries obtained from the MF and WL methods differs. According to the MF method, the region $-0.78 < J_2 /J_1 < -0.5$ exhibits first-order transitions, with continuous transitions elsewhere. In contrast, the WL algorithm identifies a broader first-order region, ranging from $-1.5<J_2/J_1 < -0.5 $, which is inconsistent with the tricritical and continuous transition scenarios predicted by the MF method.

			Details of the first-order transition identified by the WL algorithm are presented in Fig.~\ref{fig:J3=0.6}. The first column displays $U/N$ versus $T$, while the second column shows the energy distribution at the transition point. In Fig.~\ref{fig:J3=0.6} (a1), the energy shows a smooth upward slope and the single peak of $P(E/N,T_c)$ in Fig.~\ref{fig:J3=0.6} (a2) indicates a continuous phase transition at $J_2/J_1=0$. 
			In Figs.~\ref{fig:J3=0.6} (b1) and (b2), 
			although obvious energy jumps are observed and double peaks are detected for small system sizes (e.g., $L=32$), the double peaks have clearly vanished for $L=64$ and $L=96$.
			This indicates that the phase transition here remains a continuous phase transition. 
			In Figs.~\ref{fig:J3=0.6} (c1), (c2), (d1), and (d2) at $J_2/J_1=-0.75$, -1, the energies exhibit a jump as the temperature increases. This indicates a latent heat of phase transition, a hallmark of first-order phase transitions.
			The presence of a double-peak structure in the distribution confirms the conclusion.
			
			The fitted exponents $y_t$ are $1.00(3)$, $1.2(2)$, $1.9(1)$, $1.9(1)$, and $2.2(3)$, corresponding respectively to the parameter sets with $J_2/J_1 = 0$, $-0.5$, $-0.75$, $-1$, and $-1.5$ (with $J_3/J_1 = 0.6$ fixed for all cases).
			Specifically, for the first two parameters, i.e., when $J_2/J_1 = 0$ and $ -0.5$, the obtained $y_t$ values are consistent with the behavior of an Ising phase transition. For the latter three parameter sets ($J_2/J_1 = -0.75$, -1, -1.5), the $y_t$ exponents align with the characteristics of a distinct first-order phase transition. These results are consistent with the behavior indicated by the bimodal energy distributions. All results are summarized in Table~\ref{tab:0.6}.

		
		\begin{table}[tbh]
			\caption{Fitting results for the $C_V$ using the ansatz Eq.~\ref{eq:Cv} at $J_3/J_1=0.6$. The numbers in parentheses indicate errors, representing three times the standard deviation.
			}
			\begin{tabular*}{0.45\textwidth}{l@{\extracolsep{\fill}}c@{\hspace{8pt}}c@{\hspace{8pt}}l}  
				\hline\hline
				${J}_{2}/{J}_{1}$ & ${y}_{t}$ & ${T}_{c}$ & Type \\ \hline
				0       & 1.00(3) & 3.134(1) & Ising type   \\
				-0.5    & 1.2(2)  & 0.454(3) & Ising type   \\
				-0.75   & 1.9(1)  & 0.758(1) & first order  \\
				-1.0    & 1.9(1)  & 0.859(6) & first order  \\
				-1.5    & 2.2(3)  & 0.963(3) & first order  \\ \hline\hline
			\end{tabular*}
			\label{tab:0.6}
		\end{table}
		\subsubsection{$J_3/J_1=0.2$}

		\begin{figure}[h]
			\centering  \includegraphics[width=\linewidth]{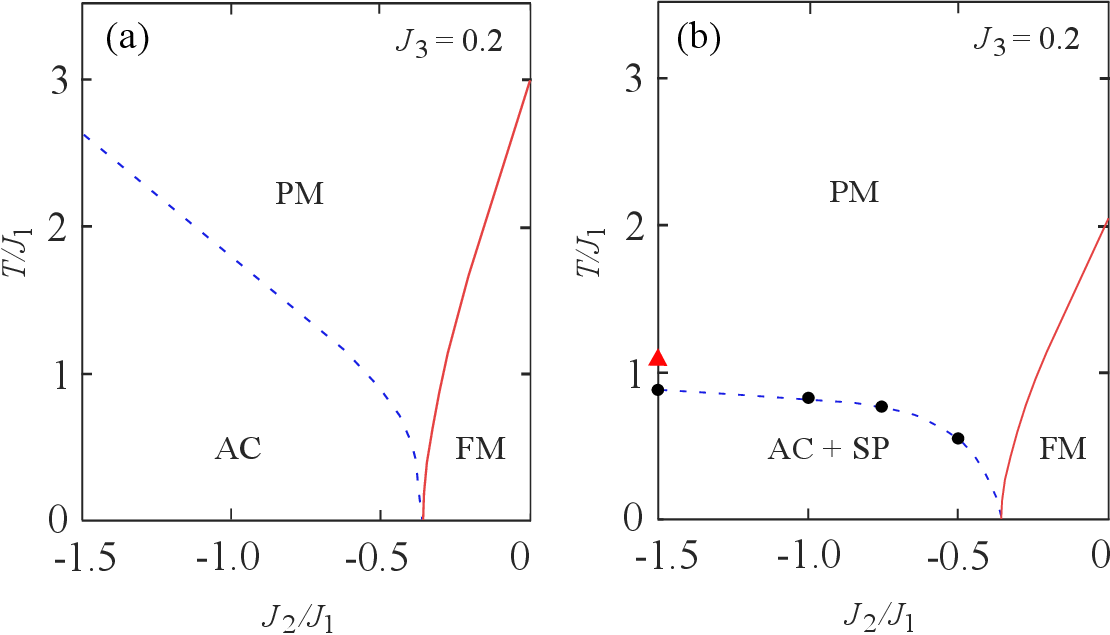}
			\caption
			{{Finite temperature phase diagram of $J_3/J_1=0.2$: (a) from Ref.~\cite{th1} by the MF method; (b) by the WL and PT methods. Solid lines indicate continuous phase transitions, whereas dashed lines represent first-order transitions. The multi-peak phenomenon in the specific heat is observed near the red-triangle-marked position at $J_2/J_1=-1.5$.}}
			
			\label{fig:phase_J3_0.2}
		\end{figure}
		
		\begin{figure}[t]
			\centering  \includegraphics[width=\linewidth]{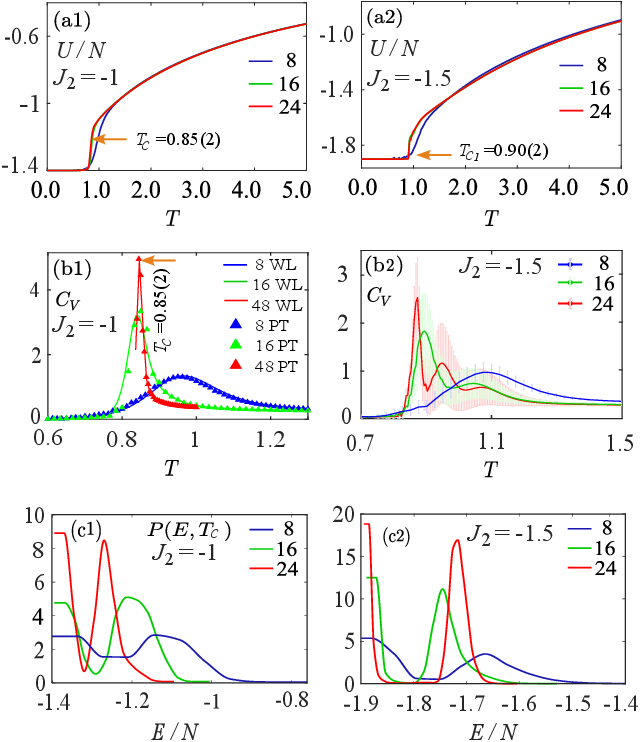}
			
			\caption{At $J_3/J_1 = 0.2$,  the temperature dependences of (a) the energy, (b) the specific heat, as well as (c) the double-peak distribution of the energy  for various system sizes. Left column: $J_2/J_1 = -1$; Right column: $J_2/J_1 = -1.5$.}
			\label{fig:d_peak_J3_0.2}
		\end{figure}
		
		\begin{figure*}[t]
			\includegraphics[width=1.0\linewidth]{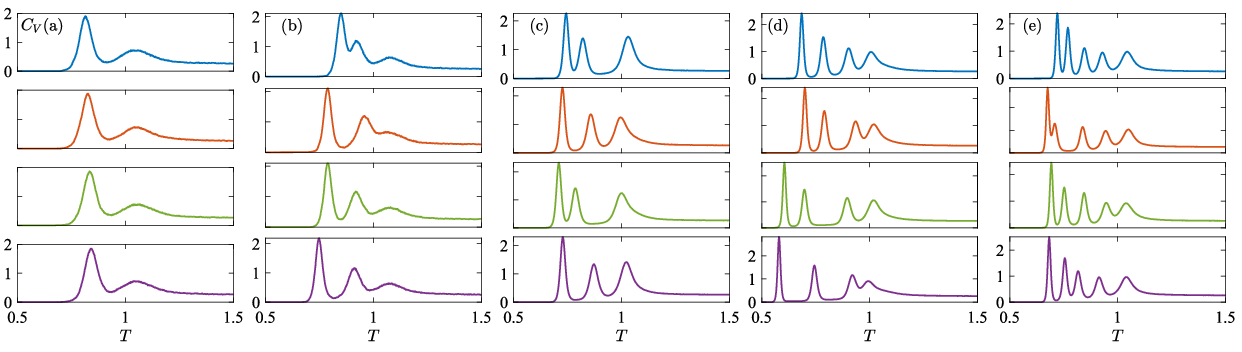}
			\caption{Comparison of specific heat obtained by the two-replica cluster updating population annealing method~\cite{keisermachta2026} for different system sizes $L=16,24,32,40,48$; the five columns correspond to these five sizes, respectively. The four curves in each column correspond to results from four different random number seeds. The number of peaks converges consistently; for $L=16,24,32,40,48$, the peak counts are 2, 3, 3, 4, and 5, respectively.}
			\label{fig:foursizesCv}
		\end{figure*}

		In this section, we fix $J_3 = 0.2$ and plot the finite-temperature phase diagram in the $J_2/J_1$-$T/J_1$ plane. In Fig.~\ref{fig:phase_J3_0.2}(a), we show the phase boundaries obtained using the MF method, while Fig.~\ref{fig:phase_J3_0.2}(b) presents the results from the PT and WL algorithms. The AC+SP phase is observed in the lower-left corner of the phase diagram, which differs from previous results where only the AC phase was identified~\cite{th1,wildes2020}.
		
		Within the range $-1 < J_2/J_1 < -0.35$, elevating the temperature induces distinct first-order phase transitions between the SP+AC phase and the paramagnetic (PM) phase. In contrast, within $-1.5 < J_2/J_1 < -1$, temperature elevation similarly triggers evident first-order phase transitions; additionally, extra peaks emerge in the specific heat curve—a signature of competing interactions in frustrated systems that underlies the first-order phase transition behavior. Notably, the number of these peaks is dependent on the lattice size: for example, lattices with $L = 8$, $16$, and $24$ exhibit 1, 2, and 3 specific-heat peaks, respectively.

		Figure~\ref{fig:d_peak_J3_0.2} illustrates the detailed characteristics of the phase transitions. Specifically, the left and right columns of the figure correspond to the cases with $J_2 = -1$ and $J_2 = -1.5$, respectively. Panels (a1) and (a2) of Fig.~\ref{fig:d_peak_J3_0.2} plot the curves of $U/N$ versus $T$, where energy jumps are observed at the critical temperatures $T_c = 0.85(2)$ and $T_c = 0.90(2)$, respectively. These energy jumps confirm the occurrence of first-order phase transitions, and the evidence for such transitions is conclusive: the bimodal energy distributions shown in Figs.~\ref{fig:d_peak_J3_0.2}(c1) and (c2) become increasingly pronounced with increasing lattice size.
		
		For $J_2 = -1$, only a single specific heat peak is observed across different lattice sizes; notably, the peak for the large lattice ($L = 48$) is located at $T_c = 0.85(2)$. In contrast, stronger frustration interactions arise for $J_2 = -1.5$, making the WL method unable to yield reliable results. Thus, we adopt the PT method to simulate the system, with thermal parameters set as follows: $5\times10^6$ thermal steps, $2\times10^4$ bins, 1000 Monte Carlo steps, and 200 temperature points distributed in the range of $0.5$ to $1.5$. Although the error bars of the specific heat remain relatively large, the multi-peak phenomenon is still distinctly observable.

		Notably, in our PT simulations, the AC configuration is adopted as the initial state for the lowest-temperature replica. With identical numbers of Monte Carlo steps, simulations started from the AC configuration yield smoother results than those with random initial configurations, since random starting states tend to remain trapped in long-lived metastable states and require vastly longer thermalization times.

		
		For Fig.~\ref{fig:d_peak_J3_0.2}(c1), we adopt the temperatures of $0.984$, $0.853$, and $0.8285$ for lattice sizes $L = 8$, $16$, and $24$, respectively, for the construction of the equal-height profile. It should be emphasized that these chosen temperatures are not required to exactly match the thermodynamic transition temperature $T_c$. For Fig.~\ref{fig:d_peak_J3_0.2}(c2), the temperatures corresponding to the same lattice sizes and equal-height criterion are $1.0234$, $0.941$, and $0.930$, respectively.
		
		\subsubsection{The multipeak phenomenon of specific heat }

		\begingroup As described in Ref.~\cite{multi_peaks}, the commonly observed multiple-peak behavior of specific heat corresponds to the Schottky anomaly~\cite{PhysRevB.46.5405,PhysRevLett.79.3451,Ramirez1994}, where the specific heat exhibits an anomalous sharp peak in addition to the conventional one. This phenomenon exists in non-frustrated systems, such as the classical spin-1 system, e.g., the Blume-Capel model~\cite{kwak2015first}.For frustrated real materials, the multi-peak Schottky effect has been experimentally observed~\cite{VILLUENDAS2016282,ITO2017390,Gondek_2007,Schobinger-Papamantellos_2008,O’Flynn_2014}.The multi-peak specific heat discussed in Ref.~\cite{multi_peaks} originates from the fact that the stepwise residual entropy makes the entropy decrease in a staircase manner with temperature, which in turn produces multiple peaks in the specific heat. Ref.~\cite{multi_peaks} found that a system with $n$ independent interactions can exhibit at most \(n+1\) peaks. However, this mechanism applies only to frustrated systems without first-order transitions as the reference emphasized~\cite{multi_peaks}; in the strongly frustrated parameter regime studied here, where first-order transitions occur, the mechanism does not apply, as evidenced by the following.
		
		In Fig.~\ref{fig:foursizesCv}, we observe multi-peak behavior in our model. Each column has  four curves, and  they correspond to different random-number seeds and are nearly identical, although the peak positions may exhibit slight fluctuations or shifts.  The specific heat exhibits three peaks at both \(L=24\) and \(L=32\), four peaks at \(L=40\), and five peaks at \(L=48\). Although \(L=24\) and \(L=32\) both show three peaks, the height and sharpness of the last peak differ with system size (e.g., the last peak at \(L=32\) is noticeably higher and sharper than that at \(L=24\)).
		Notably, if one naively expects the number of transition peaks to be bounded by \(n+1=4\) from the three independent couplings (\(J_1\), \(J_2\), \(J_3\)), the number of peaks at our largest size (five for \(L=48\)) exceeds this bound. 
		
		In this strongly frustrated regime, neither a pure Metropolis scheme combined with parallel tempering nor the pure Wang–Landau algorithm yielded reliable results for larger sizes (e.g., $L\ge32$). Fortunately, using the two-replica cluster algorithm combined with population annealing by Keiser and Machta~\cite{keisermachta2026}, we are able to reproduce the multi-peak specific heat precisely for \(L=8,16,24,32,40,48\) within reasonable computational time. The typical parameters are listed in Table~\ref{tab:pa_params}.
		
		The multi-peak phenomenon and its size dependence observed here are reminiscent of a recent high-resolution simulation of the two-dimensional ANNNI model~\cite{keisermachta2026}: there, too, the number of sharp specific-heat peaks increases with system size, and these peaks correspond to first-order-like transitions between states with different numbers of defect lines.
		The multi-peak behavior we observe may correspond to transitions or melting between ordered states with different wave vectors characterized by  structure factors.
		Earlier transfer-matrix studies suggested that multiple peaks eventually merge into a single peak; this conclusion originates from a fundamental limitation of the transfer-matrix method: it effectively probes a quasi-one-dimensional geometry rather than the true two-dimensional system~\cite{multi_peak_mc,PhysRevB.103.094441,PhysRevB.104.144429}.
		\endgroup
		
		\begin{table}[t]
			\caption{Simulation parameters of the population annealing runs. Here $\delta\beta_1$ and $\delta\beta_2$ are the inverse-temperature increments outside and inside the fine-scan window, respectively; $R$ is the number of replicas, $S$ the number of two replica  cluster (TR) sweeps per temperature step, and $n_{\rm met}$ the number of Metropolis sweeps per replica per temperature step.}
			\label{tab:pa_params}
			\begin{tabular}{cccccc}
				\hline\hline
				~$L$~~ & ~~$\delta\beta_1$~~ & ~~$\delta\beta_2$~~ & $R$~~ & ~~$S$~~ & $n_\text{metropolis}$ \\ \hline
				8  & 0.001 & 0.002 & 4000 & 20 & 4 \\
				16 & 0.001 & 0.002 & 4000  & 20 & 4 \\
				24 & 0.001 & 0.002 & 4000  & 20 & 4 \\
				32 & 0.0002 & 0.0005 & 50000 & 10 & 5 \\
				40 & 0.002 & 0.005 & 100000 & 10 & 5 \\ 
				48 & 0.002 & 0.005 & 100000 & 10 & 5 \\\hline\hline
			\end{tabular}
		\end{table}
		{Following the protocol summarized in Table~\ref{tab:pa_params}, the run cools along an inverse-temperature ladder from high to low temperature: \(\beta\) ranges from \(0.20\) to \(3.33\), corresponding to a cooling from \(T=5\) down to \(T=0.3\). To resolve the transition region, the temperature range is divided into three intervals: the high-temperature interval (\(\beta<0.625\)) and the low-temperature interval (\(\beta>1.25\)) use a coarser step of $\delta\beta_2$, while the peak interval (\(0.625\le\beta\le1.25\), i.e. \(T\in[0.8,1.6]\)) uses a finer step of $\delta\beta_1$ so that the specific-heat peak at \(T\approx1.0\)–\(1.1\) is well resolved.}s
		
		\section{Conclusion and discussion}
		\label{sec:con}
		
		In this work, we systematically reexamine the classical $J_1$-$J_2$-$J_3$ model on the honeycomb lattice via the WL algorithm, delivering new insights into the system’s properties. We first clarify the degeneracy of the AC phase: contrary to prior reports of 4- or 12-fold degeneracy~\cite{zigzag,th1}, the AC phase exhibits 20-fold degeneracy, with 16 newly identified degenerate states corresponding to spiral configurations, which can be characterized by 4 bright spots in the momentum-space structure factor. 
		
		Regarding phase transitions, our WL results reveal distinct behaviors across parameter regimes: for $J_3/J_1 = 1.2$, the stripe (ST)-paramagnetic (PM) transition hosts a tricritical point near $J_2/J_1 \approx -0.75$,
		while the determination of more precise values requires extensive calculations and analyses~\cite{PhysRevE.78.061104,Tri_cons_2d,Tri_const_3d,Tri_Edge_phas_tran,Tri_geomet_prop_2d,Tri_red_bond,Tri_field-driven_blume-capel}, which fall beyond the scope of this paper.
		For $J_3/J_1 = 0.6$, the ST-PM transition is either continuous or first-order; when $J_3/J_1 = 0.2$, the AC+SP-PM transition is also first order. In the strong frustration regime ($J_2/J_1 = -1.5$), we further observe multi-peak behavior in the specific heat~\cite{multi_peak_mc,PhysRevB.103.094441}. When $J_3 = 0$, our findings align with previous Metropolis and parallel tempering studies of the $J_1$-$J_2$ honeycomb model~\cite{ising}, confirming the consistency of our approach.
		
		This work establishes a foundation for further exploration of the $J_1$-$J_2$-$J_3$ model. Future investigations will focus on higher-precision calculations of phase transition points and critical exponents, leveraging advanced methods such as tensor network Monte Carlo~\cite{tnmc2,chen2025,chen2025twodimensional,tnmc}. We also aim to extend the framework: incorporating external magnetic fields to map field-dependent phase diagrams, expanding parameter ranges to stronger frustration regimes, investigating high-spin generalizations~\cite{wildes2020}, and extending the analysis to the quantum transverse-field $J_1$-$J_2$-$J_3$ Ising model, for exploring quantum phase transitions that complement the classical thermal transitions studied here.\\

		{\it Acknowledgements:} We thank Yi Hu and Patrick Charbonneau for helpful discussions. W. Z. was supported by the Hefei National Research Center for Physical Sciences at the Microscale (Grant No. KF2021002) and the Fundamental Research Program of Shanxi Province (Grant Nos. 202303021221029).
		Y. J. was supported by the National Natural Science Foundation of China (Grant
		No. 12275263) and the Natural Science Foundation of Fujian Province of China (Grant No. 2023J02032).

		
		\appendix

		\begin{figure}[t]
			\centering  \includegraphics[width=0.8\linewidth]{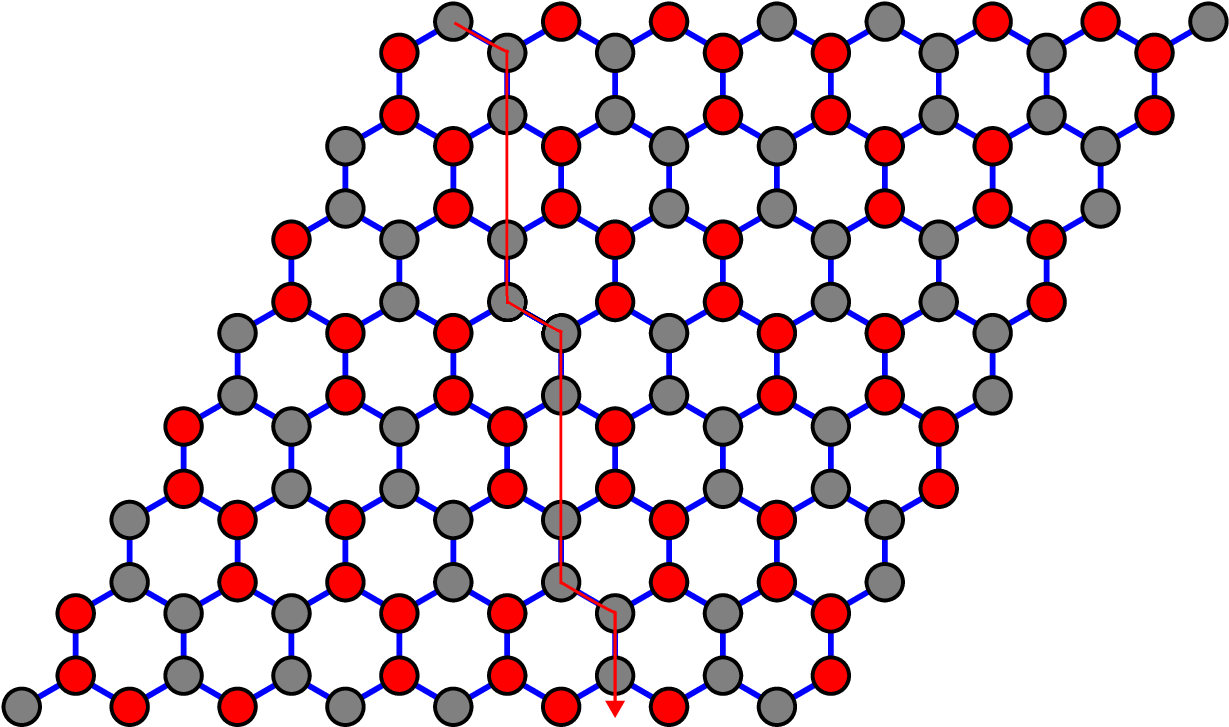}
			\caption
			{A typical configuration in the SP phase on the hexagonal lattice; arrows indicate the spiral direction.}
			\label{spiral_honeycomb}
		\end{figure}
		\begin{figure}[t]
			\centering  \includegraphics[width=\linewidth]{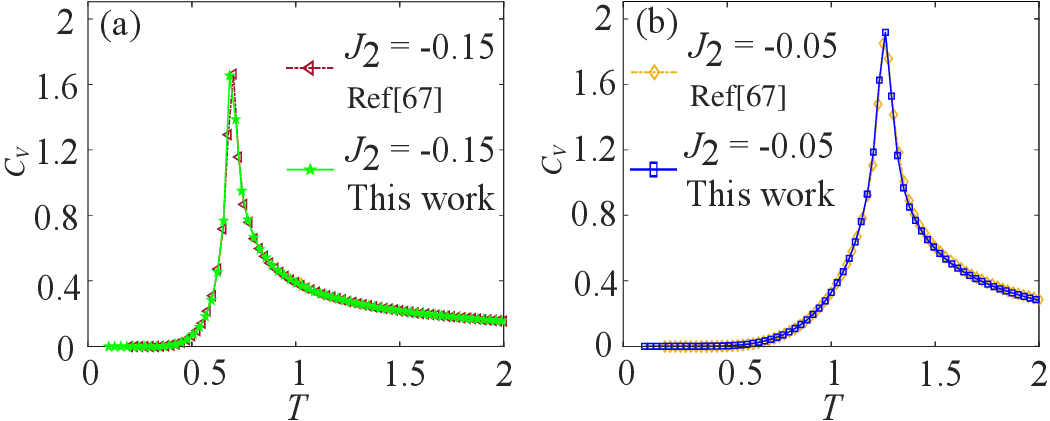}
			\caption
			{Specific heat  as a function of temperature for $J_2 = -0.15$ and $-0.05$, comparing our results with those from Ref.~\cite{ising}.}
			\label{comp_liter}
		\end{figure}
		\begin{figure}[b]
			\centering
			\includegraphics[width=\linewidth]{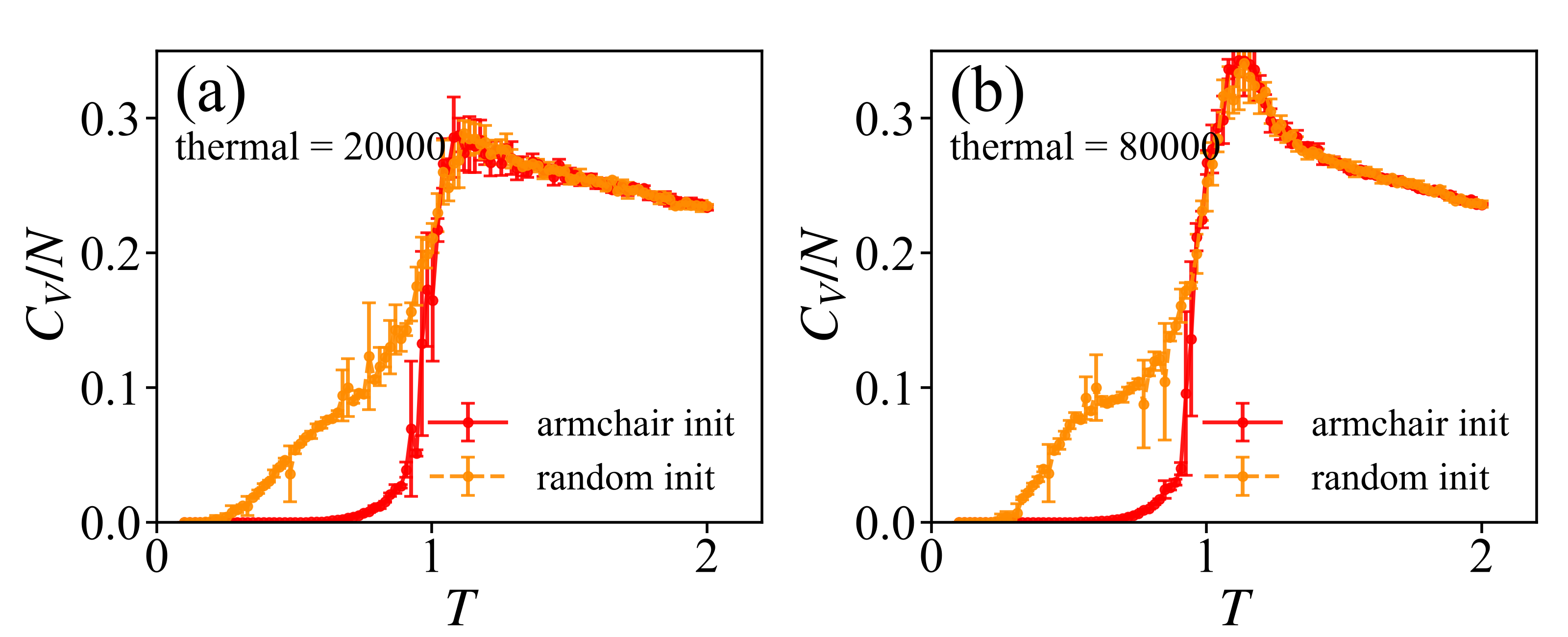}
			\caption{Specific heat $C_V/N$ as a function of temperature $T$ for $L=16$ at $J_2/J_1=-1.5$ and $J_3/J_1=0.2$, computed with PT initialized from the AC ground state (red) and from random configurations (orange): (a) thermal$=20000$ and (b) thermal$=80000$ Monte Carlo steps. Increasing thermalization steps does not pull the random-initialized results away from the metastable regime.}
			\label{fig:cvs_low_med}
		\end{figure}
		\section{A snapshot in the real honeycomb lattices}
		\label{sec:app}
		In the main text, to facilitate the observation of symmetries of the configuration such as translational symmetry and mirror reflection symmetry, we compress the hexagonal lattice into a brick-wall lattice for investigation. In this appendix, we directly place a specific SP configuration on the hexagonal lattice for easy reference.

		\section{Comparison of the data with Ref.~\cite{ising}}
		
		We present in Fig.~\ref{comp_liter} the temperature dependence of the specific heat for $J_3=0$ at $J_2=-0.15$ and $-0.05$, and compare our results with those reported in Ref.~\cite{ising}. As shown in Fig.~\ref{comp_liter}, our data are in very good agreement with the literature results over the whole temperature range. In particular, the overall behavior and the location of the characteristic features are consistent with those obtained in Ref.~\cite{ising}. This agreement confirms the reliability and correctness of our simulation approach for the $J_3=0$ case.

		Effect of the initial configuration in parallel tempering
		\label{sec:PT_init}
		
		In the parallel tempering (PT) simulations reported in the main text, the AC ground-state configuration is adopted as the initial state of the lowest-temperature replica. Here we justify this choice. We consider the strongly frustrated regime $J_2/J_1=-1.5$ and $J_3/J_1=0.2$ (the same parameters as in Fig.~7 of the main text) and compare PT runs initialized from the AC ground state (red) with those initialized from random spin configurations (orange), for $L=16$.
		
		Figure~\ref{fig:cvs_low_med} shows the specific heat $C_V/N$ as a function of temperature $T$ at two thermalization lengths. With a short thermalization [panel (a), thermal$=20000$ Monte Carlo steps], the random initialization remains trapped in long-lived metastable stripe or domain-wall states: its low-temperature specific heat is strongly suppressed and exhibits large error bars, whereas the AC initialization already yields a smooth curve. Increasing the thermalization to $80000$ steps [panel (b)], the random-initialized results still remain in the metastable regime and do not converge to those initialized from the AC state.
		
		Since the observables reported in the main text are energy and specific heat, and all degenerate ground states share identical energies, the low-temperature specific heat is controlled by excitations above the ground-state manifold rather than by the particular starting sector. Starting from the ordered AC state merely brings the lowest-temperature replica close to equilibrium and bypasses the slow thermalization barrier associated with metastable states.

		\bibliography{ref.bib}
		
	\end{document}